\documentclass[article]{aastex631}
\usepackage{subfigure}
\graphicspath{{./}{figures/}}

\begin{document}

\title{An Asymptotic Matching Method for Analyzing Non-Planar Orbits in Disk Galaxies}

\author[0009-0003-3513-1887]{Sena Ghobadi}
\affiliation{School of Physics and Center for Relativistic Astrophysics, 837 State St NW, Georgia Institute of Technology, Atlanta, GA 30332, USA; \href{mailto:sghobadi6@gatech.edu}{sghobadi6@gatech.edu}}



\begin{abstract}
Modeling the orbital dynamics of objects in galactic disks is crucial to understanding the stability and evolution of disk galaxies. While studies of galactic orbits are largely dominated by $N$-body simulations, perturbative analytical models offer a computationally inexpensive and conceptually  insightful way of analyzing galactic dynamics. We utilize perturbation theory and the method of matched asymptotics to develop a technique by which the vertical motion of a point mass perpendicular to a thin axisymmetric disk galaxy can be computed analytically to high precision. The objective of this study is to provide an accurate model of non-planar dynamics that could be employed in galactic simulations to bypass computationally expensive integration. We construct a general solution for the dynamics at small $z$ displacements from the plane of the disk using perturbation theory. We solve for the dynamics at very large displacements from the plane of the disk and construct a function that interpolates the solutions at each length scale. To demonstrate the method's utility, we apply it to two commonly used models for the vertical structure of a galactic disk, the exponential model and the isothermal model. Finally, using the principle of adiabatic invariance, we analyze how to relate variations in radial motion to variations in the vertical motion and discuss possible applications of the model in numerical astrophysics.

\end{abstract}


\section{Introduction}\label{sec:intro}
Orbital dynamics play a central role in our understanding of galaxies, from the motion of stars and gas in disks to the formation of dark matter structures. The study of vertical motion in galactic disks, in particular, is essential for probing the structure and stability of disk galaxies, as well as for determining the distribution of baryonic and dark matter within them \citep{2013ApJ...779..115B, 2014JPhG...41f3101R}. These vertical oscillations influence the thickening of galactic disks and the kinematics of stars, both of which are critical for understanding galaxy evolution and secular processes such as bar formation or spiral arm dynamics \citep{2014RvMP...86....1S}.
\newline
\indent The calculation of orbits is dominated by numerical methods since the equations of motion of most systems in stellar and galactic dynamics are comprised of nonlinear ordinary differential equations (ODEs) \citep{binney2008galactic}. The complexity of the equations of motion is dictated by the form of the Newtonian gravitational potential $\Phi(\vec{r})$ which is too complicated to be expressed analytically for many disk mass distributions. Few systems exist for which $\Phi(\vec{r})$ is simple enough that the resulting ODEs have analytical solutions. An alternative method of analysis exists in perturbation theory which can be used to analyze the stability of stellar systems while providing intuitive insight into the mechanisms of orbits. We propose a method by which the motion of objects perpendicular to the disk in galaxies with a smooth gravitational potential can be computed analytically to reasonably high precision using the concept of matched asymptotic expansions. The method of matched asymptotic expansions is primarily utilized in the theory of fluids where solutions vary significantly at different scales due to the presence of boundary layers \citep{1959flme.book.....L}. We present similar arguments demonstrating that the vertical motion of objects in a galactic disk is easily calculated using analytic methods at small and large length scales. Asymptotic matching is then used to smoothly interpolate the solutions at the two distinct length scales so that the dynamics may be accurately computed at all possible length scales.
\newline
\indent The paper is organized as follows. In Section \ref{sec:methods}, we derive the analytical model using the machinery of action-angle variables and classical perturbation theory. In particular, we begin with a basic analysis of small displacements using the epicycle approximation and then generalize the results using a perturbative expansion. We similarly derive the dynamics at large displacements outside the disk and introduce the matching procedure to be used in constructing the full solution. In Section \ref{sec:apps}, we illustrate the utility of the matching method by applying it to two well-studied vertical galactic structures, the exponential disk and the isothermal disk. We include numerical tests of the model to reinforce its effectiveness. Finally, we discuss the implications of the model in Section \ref{sec:discussion} and review the results and possible numerical applications in Section \ref{conc}.
\section{Methods}
\label{sec:methods}
\subsection{Dynamics of Small Oscillations} \label{sub:2.1}
We consider the general Hamiltonian of a point mass orbiting in an axisymmetric potential in Equation (\ref{eq:1}) working in a cylindrical coordinate system with the center of the disk acting as the origin.  

\begin{equation} \label{eq:1}
    \mathcal{H} = \frac{\vec{p}^2}{2} + \Phi(r,z)
\end{equation}
We have $\vec{p}$ representing the momentum of the particle and $\Phi(r,z)$ representing the galactic potential. The potential may be computed as a solution to the Poisson equation. For an axisymmetric potential, the Poisson equation in cylindrical coordinates is given by Equation (\ref{eq:2}). 

\begin{equation} \label{eq:2}
    \frac{1}{r}\frac{\partial}{\partial r}\left(r \frac{\partial \Phi}{\partial r}\right) + \frac{\partial^2 \Phi}{\partial z^2} =  4\pi G \rho(r,z)
\end{equation}
The potential is governed by the galaxy's axisymmetric density distribution $\rho(r,z)$. Our goal is to derive approximate solutions for the non-planar dynamics of systems governed by Equations (\ref{eq:1}) and (\ref{eq:2}) in thin disk galaxies. A thin disk galaxy is characterized by its flatness. We define the flatness condition to be $R_g \gg z_g$ where $R_g$ represents the scale radius of the galaxy and $z_g$ represents its scale height. We are interested in bound, quasi-circular orbits typical for stars in a galaxy where the average radial oscillations of each orbit are much smaller than the scale radius of the galaxy. For now, we impose the assumption that the radius of the object's orbit is fixed. This implies that the radial component of the momentum is constant and that the potential is a function of the $z$-coordinate only. We can then decouple $r$ and $z$ and isolate our focus to the Hamiltonian governing the $z$-dynamics given in Equation (\ref{eq:3}).
\begin{equation}\label{eq:3}
    \mathcal{H}_z = \frac{p_z^2}{2} + \Phi(z)
\end{equation}
Note that $p_z$ is the vertical momentum of the object and the potential is now an explicit function of $z$ only. The challenge of analyzing this orbit arises from finding an expression for $\Phi(z)$. By the symmetry of our system, we reason that for any orbit at fixed $R$, there must be a potential minimum in the plane of the disk at $z = 0$. Therefore, as a first step in understanding the dynamics, we will expand the potential about $z = 0$ and attempt to model the oscillatory behavior of the mass at small amplitudes $z_0$ from the disk plane using Equation (\ref{eq:4}). 
\begin{equation}\label{eq:4}
    \mathcal{H}_z \approx \frac{p_z^2}{2} +\frac{1}{2}\Phi''(0) z^2
\end{equation}
Note that primes denote differentiation with respect to $z$. Of course, this Hamiltonian corresponds to simple harmonic motion as would be expected for any system near a potential minimum. The goal becomes evaluating the natural frequency $\omega_0$ given by the relation in Equation (\ref{eq:5}).
\begin{equation} \label{eq:5}
\omega_0^2 = \Phi''(0)
\end{equation}
Without knowing the potential explicitly, the only way to evaluate the natural frequency is by solving for the second derivative term in Equation (\ref{eq:2}). For a flattened disk system, we argue that the radial term in Equation (\ref{eq:2}) contributes negligibly \citep{binney2008galactic}. This follows from the fact that as the density of the disk becomes more concentrated around $z=0$, the potential becomes much steeper about $z=0$ which decreases the relative contribution of the radial term in Equation (\ref{eq:2}). Thus, we make the following approximation in Equation (\ref{eq:6}).
\begin{equation} \label{eq:6}
    \omega_0^2 \simeq 4\pi G \rho(r, 0)
\end{equation}
For thicker disks, the radial term cannot be neglected, although the object still undergoes simple harmonic oscillations. Therefore, at small oscillation amplitudes, $z_0$, where small amplitude is defined by the condition $z_0 \ll z_g$, we obtain sinusoidal oscillations with frequency $\omega_0$ given by Equation (\ref{eq:6}).
\subsection{Perturbative Expansion and Corrections}\label{sub:2.2}
To analytically model the system's dynamics for larger amplitudes $z_0 \sim z_g$, we require perturbation theory. We may obtain increasingly accurate solutions by expanding the Hamiltonian to higher order terms in $z$. To obtain the first order correction, we expand the potential to third order in $z$ in Equation (\ref{eq:7}).

\begin{equation} \label{eq:7}
    \mathcal{H}_z = \frac{p_z^2}{2} +\frac{1}{2}\omega_0^2 z^2+ \frac{1}{6}\phi_3 |z|^3
\end{equation}
We have introduced the notation $\phi_n \equiv \Phi^{(n)}(0)$ for convenience. The absolute value of the cubic term is taken due to the reflectional symmetry of the system which constrains the Hamiltonian to be even with respect to $z$. We treat the cubic term as the perturbing Hamiltonian $\mathcal{H}_1$. Following the techniques of canonical perturbation theory \citep{goldstein1980classical}, we express the perturbing Hamiltonian using the unperturbed solution written in terms of action-angle variables. The unperturbed solution is given by Equation (\ref{eq:8}).
\begin{equation} \label{eq:8}
    z = \sqrt{\frac{J}{\pi \omega_0}} \sin{(\omega_0 t + 2\pi \beta)}
\end{equation}
Where $J$ and $\beta$ are the action and angle variables respectively. Therefore, the perturbing Hamiltonian is given by Equation (\ref{eq:9}). 

\begin{equation} \label{eq:9}
    \mathcal{H}_1 = \frac{1}{6} \phi_3 \left(\frac{J}{\pi \omega_0}\right)^{3/2} \left|\sin^3{(\omega_0 t + 2\pi \beta)}\right|
\end{equation}
Hamilton's equations yield the resulting time dependencies of $\beta$ and $J$ in Equations (\ref{eq:10}) and (\ref{eq:11}). 
\begin{equation} \label{eq:10}
    \dot{\beta} = \frac{\partial  \mathcal{H}_1}{\partial J}
\end{equation}
\begin{equation} \label{eq:11}
    \dot{J} = -\frac{\partial  \mathcal{H}_1}{\partial \beta}
\end{equation}
We want to compute the secular contributions $\langle \dot{J} \rangle$ and $\langle \dot{\beta}\rangle$ to the solution where $\langle \rangle$ indicates a time-average over a period of the motion. It is easy to compute that $\langle \dot{J} \rangle$ = 0 which is intuitive since there is no dissipation in the system. Evaluating $\langle \dot{\beta}\rangle$ and integrating with respect to time yields the secular contribution in Equation (\ref{eq:12}).
\begin{equation} \label{eq:12}
    \langle \beta\rangle = \frac{\phi_3}{3\pi^2 \omega_0} z_0 t
\end{equation}
We have used the fact that $J = \pi z_0^2 \omega_0$ for a simple harmonic oscillator with amplitude $z_0$. The first-order corrected frequency $\omega_1$ is then given by Equation (\ref{eq:13}).
\begin{equation} \label{eq:13}
    \omega_1 = \omega_0 \left(1 + \frac{2\phi_3}{3\pi \omega_0^2} z_0\right)
\end{equation}
There is a small perturbation parameter $\varepsilon$ hidden in this expression that we can identify using dimensional analysis. The $\phi_3$ coefficient stems from the third order derivative of the potential, which has units of squared frequency divided by length. The only relevant length scale in our problem is the scale height of the disk $z_g$. Thus, the relevant perturbation parameter is $\varepsilon \equiv \frac{z_0}{z_g}$. Subsequent corrections should then be higher order in $\varepsilon$.
There are many potentials for which $\phi_3 = 0$ due to evenness symmetry, so it is worth calculating the second correction with $\phi_4$. We expand the potential to fourth order in $z$ which yields the Hamiltonian in Equation (\ref{eq:14}).
\begin{equation} \label{eq:14}
    \mathcal{H}_z = \frac{p_z^2}{2} +\frac{1}{2}\omega_0^2 z^2+ \frac{1}{6}\phi_3 |z|^3 + \frac{1}{24}\phi_4 z^4
\end{equation}
Following the theory outlined in \cite{goldstein1980classical}, we can compute the second correction by evaluating $\frac{\partial \langle \eta_2 \rangle}{\partial J}$ where $\eta_2$ is defined in Equation (\ref{eq:15}).
\begin{eqnarray} \label{eq:15}
    \eta_2 = \mathcal{H}_2 + \left( 2\pi \frac{\langle \mathcal{H}_1 \rangle - \mathcal{H}_1}{\omega_0}\right) \frac{\partial \mathcal{H}_1}{\partial J}  + \frac{1}{2} \left(2\pi \frac{\langle \mathcal{H}_1 \rangle - \mathcal{H}_1}{\omega_0} \right)^2 \frac{\partial^2 \mathcal{H}_0}{\partial J^2}
\end{eqnarray}
The unperturbed Hamiltonian is simply $\mathcal{H}_0 = \frac{J\omega_0}{2\pi}$ which means the rightmost term of Equation (\ref{eq:15}) is 0. In this case, the second order perturbing Hamiltonian $\mathcal{H}_2$ is the $z^4$ term in the expansion of the potential. Time-averaging and differentiating, we obtain the following second order correction to the frequency $\omega_2$ in Equation (\ref{eq:16}).

\begin{eqnarray} \label{eq:16}
    \omega_2 = \omega_0\left[1 + \frac{2\phi_3}{3\pi \omega_0^2} z_0 + \frac{4}{\omega_0^2}\left(\frac{\phi_4}{64} +\left(\frac{4}{27\pi^2} - \frac{5}{192}\right)\frac{\phi_3^2  }{\omega_0^2} \right)z_0^2 \right]
\end{eqnarray}
As expected, we obtain a correction that is second order in $z_0$ corresponding to the $\varepsilon^2$ term in the perturbation series. As a quick verification, we can test a pendulum potential $\Phi(z) = -z_g^2 \omega_0^2 \cos{\left( \frac{z}{z_g}\right)}$ since the perturbative corrections to this potential are known. For this potential, $\phi_3 = 0$ and $\phi_4 = -\frac{\omega_0^2}{z_g^2}$. Plugging these into Equation (\ref{eq:16}) and using the definition of $\varepsilon$, we recover the expected amplitude dependence of the frequency in Equation (\ref{eq:17}).
\begin{equation}\label{eq:17}
    \omega_{pendulum} = \omega_0\left(1 - \frac{1}{16}\varepsilon^2\right)
\end{equation}
\subsection{Asymptotic Solution} \label{sub:2.3}
For the second component of our model, we consider motion outside the plane of the disk $z_0 \gg z_g$. We argue that in this regime, the motion of the mass is effectively governed by a potential that is linear in $z$ and symmetric about $z=0$. This is intuitively plausible since the field of a uniform sheet is constant. The argument follows from applying Gauss' Law to a small section of the galactic disk in Equation (\ref{eq:18}).
\begin{equation} \label{eq:18}
    \int \nabla \Phi \cdot d\vec{A} = 4\pi G \int \rho(\vec{r}) \, dV
\end{equation}
We are interested in the dynamics far away from the disk, so we construct a narrow symmetric Gaussian box with height $2z_0 \gg z_g$ centered at $z=0$ with small cross-sectional area $A$ such that the surface mass density of the disk $\Sigma$ is essentially constant over the surface. Therefore, Gauss' Law yields the field in the $z$-direction in Equation (\ref{eq:19}).

\begin{equation} \label{eq:19}
    \frac{\partial \Phi}{\partial z} = 2\pi G \int_{-z_0}^{z_0} \rho(z) \, dz
\end{equation}
For large amplitude $z_0 \gg z_g$, we can approximately evaluate Equation (\ref{eq:19}) using Equation (\ref{eq:20}).

\begin{equation} \label{eq:20}
    \frac{\partial \Phi}{\partial z} = 2\pi G \int_{-\infty}^{\infty} \rho(z) \, dz = 2\pi G \Sigma
\end{equation}
Therefore, the object undergoes an acceleration $\vec{g} = -\textrm{sgn}(z) 2\pi G \Sigma \hat{z}$ far from the disk. Though the acceleration itself is a function of $z$ and changes as the object approaches the disk, the object passes through the disk quickly, implying that the object will spend most of its vertical orbit near the extremes $z_0$ and $-z_0$. Therefore, the acceleration we computed holds for approximately the whole orbit. The potential corresponding to this type of motion is given in Equation (\ref{eq:21}) where we have defined $g \equiv 2\pi G \Sigma$.
\begin{equation} \label{eq:21}
    \Phi_z = g |z|
\end{equation}
The frequency of the motion as a function of energy or amplitude is simple to calculate for a linear potential. We will evaluate it using action-angle variables. We first note that our potential is time independent, so the Hamiltonian of the system is equal to its energy $E$ which is a constant. The energy is given in Equation (\ref{eq:22}).
\begin{equation} \label{eq:22}
    E = \mathcal{H}_z = \frac{p_z^2}{2} + g|z|
\end{equation}
For a given amplitude $z_0$, the energy of the system must be $E = gz_0$. We evaluate the action in Equation (\ref{eq:23}).
\begin{eqnarray}\label{eq:23}
    J &=& \oint p_z \ dz \nonumber \\
    &=& 4 \sqrt{2} \int_0^{z_0} \sqrt{E - gz} \, dz \nonumber \\
    &=& \frac{8\sqrt{2}}{3g} E^{3/2}
\end{eqnarray}
Finally, we compute the frequency $\omega_l$ of the motion as a function of the amplitude in Equation (\ref{eq:24}).
\begin{equation} \label{eq:24}
    \omega_l = 2\pi \frac{\partial E}{\partial J} = \frac{2\pi g}{\sqrt{32E}} = 2\pi \sqrt{\frac{g}{32z_0}}
\end{equation}
We find that the frequency scales as $z_0^{-1/2}$ which implies that the period of the motion scales as $\sqrt{z_0}$ as expected for a linear potential. One would obtain the same time scaling for a ball dropped at a height $z_0$ in a constant gravitational field.

\subsection{Matched Asymptotic Model} \label{sub:2.4}
We now combine the results of the two previous sections to construct a matched asymptotic model of the oscillation frequency $\omega_m$ for all values of $\varepsilon$ and their corresponding energies. Recall that $\varepsilon$ is the amplitude parameter $\frac{z_0}{z_g}$ that was originally used as the small perturbation parameter in Section \ref{sub:2.2}. The Taylor expansion of our matched asymptotic model should align with the perturbation expansion we calculated earlier in the limit of small $\varepsilon$. Likewise, in the limit of large $\varepsilon$, our solution should match Equation (\ref{eq:24}). We propose that the matched asymptotic solution can be expressed in the form given in Equation (\ref{eq:25}).
\begin{equation} \label{eq:25}
    \omega_m(\varepsilon) = \omega_0 \left(f(\varepsilon) + \frac{8z_g \omega_0^2 \varepsilon}{\pi^2 g}\right)^{-1/2}
\end{equation}
The function $f(\varepsilon)$ is some function of the amplitude parameter that decays to 0 for large $\varepsilon$. In principle, there are an infinite number of choices for $f(\varepsilon)$, but we can sensibly choose to use exponentials as the basis for $f(\varepsilon)$ and postulate a solution of the form given in Equation (\ref{eq:26}) with parameters $A$, $B$, and $C$ to be fit using the perturbation series.

\begin{equation}\label{eq:26}
    \omega_m(\varepsilon) = \omega_0 \left(A + Be^{-C\varepsilon} + \frac{8z_g \omega_0^2 \varepsilon}{\pi^2 g}\right)^{-1/2}
\end{equation}
We elect to use this function since it exhibits the behavior we derived in the previous sections for small and large $\varepsilon$. This is easy to see for the case of large $\varepsilon$ where the frequency is equivalent to the result in Equation (\ref{eq:24}). The most appropriate parameters can be solved for by Taylor expanding Equation (\ref{eq:25}) to second order in $\varepsilon$. The goal is then to match each coefficient with its corresponding perturbation coefficient computed using Equation (\ref{eq:16}). Equation (\ref{eq:16}) depends on the specific form of the potential, so we will demonstrate the method by applying it to two examples corresponding to commonly referenced vertical density distributions of galactic disks.

\section{Model Applications}\label{sec:apps}
\subsection{The Exponential Disk} \label{sub:3.1}
\begin{figure} 
    \centering
    
    \subfigure{
        \includegraphics[width=0.48\textwidth]{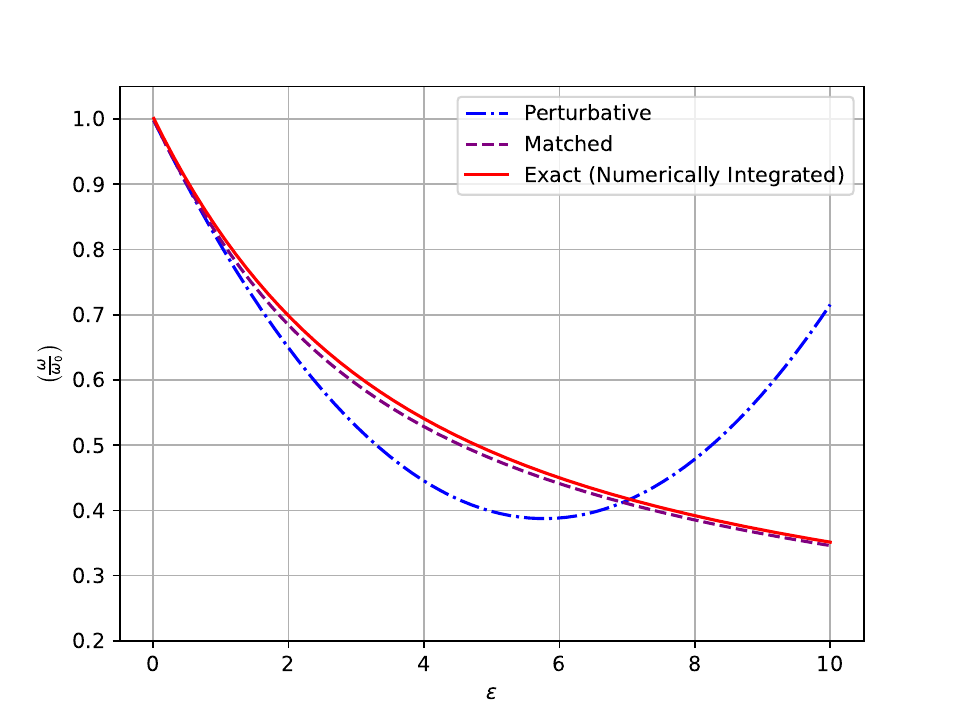}  
    }
    \subfigure{
        \includegraphics[width=0.48\textwidth]{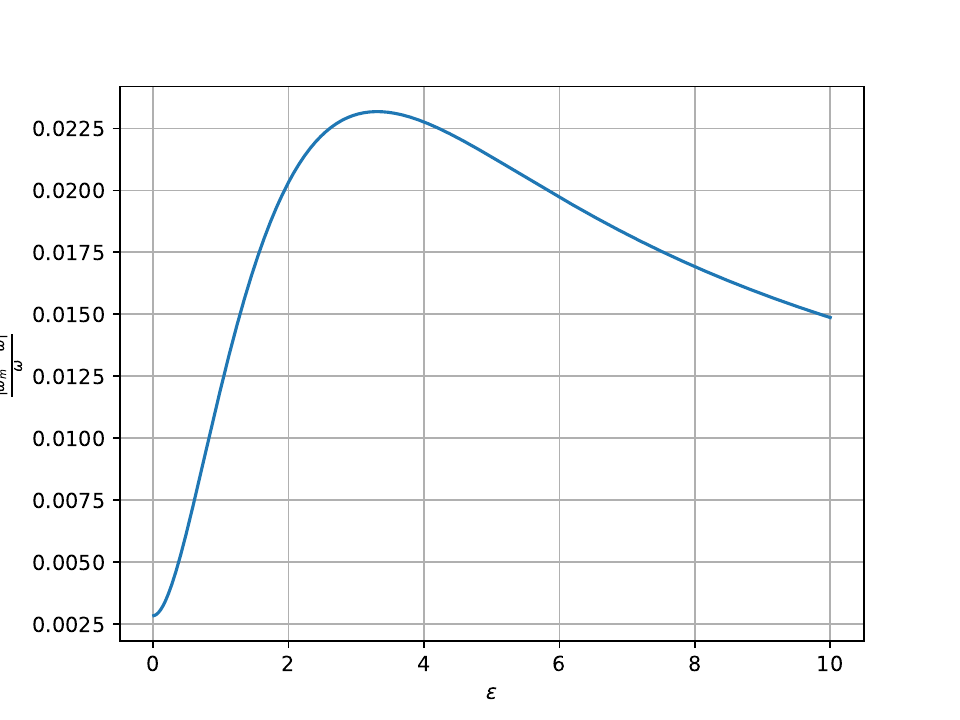}  
    }
    \caption{\textit{Left:} Direct comparison of the frequencies $\omega$, $\omega_m$, and $\omega_2$ as a function of the amplitude parameter $\varepsilon$. \textit{Right:} Relative error between the matched asymptotic solution and the numerically integrated solution which peaks at roughly 2\%. } \label{fig1}
\end{figure}

\begin{figure} 
    \centering
    
    \subfigure{
        \includegraphics[width=0.48\textwidth]{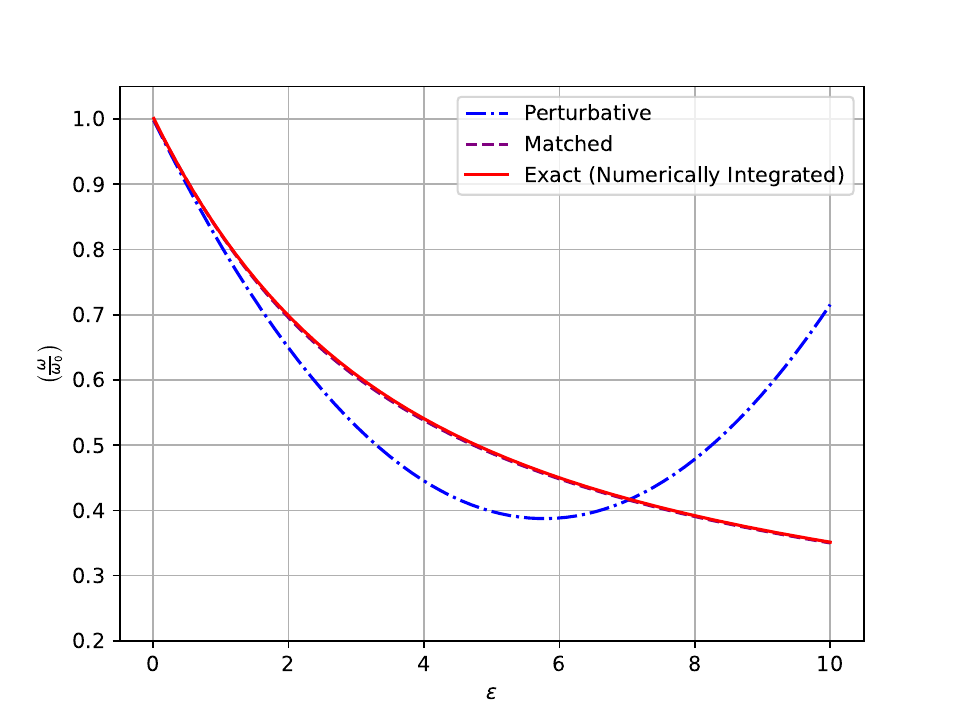}  
    }
    \subfigure{
        \includegraphics[width=0.48\textwidth]{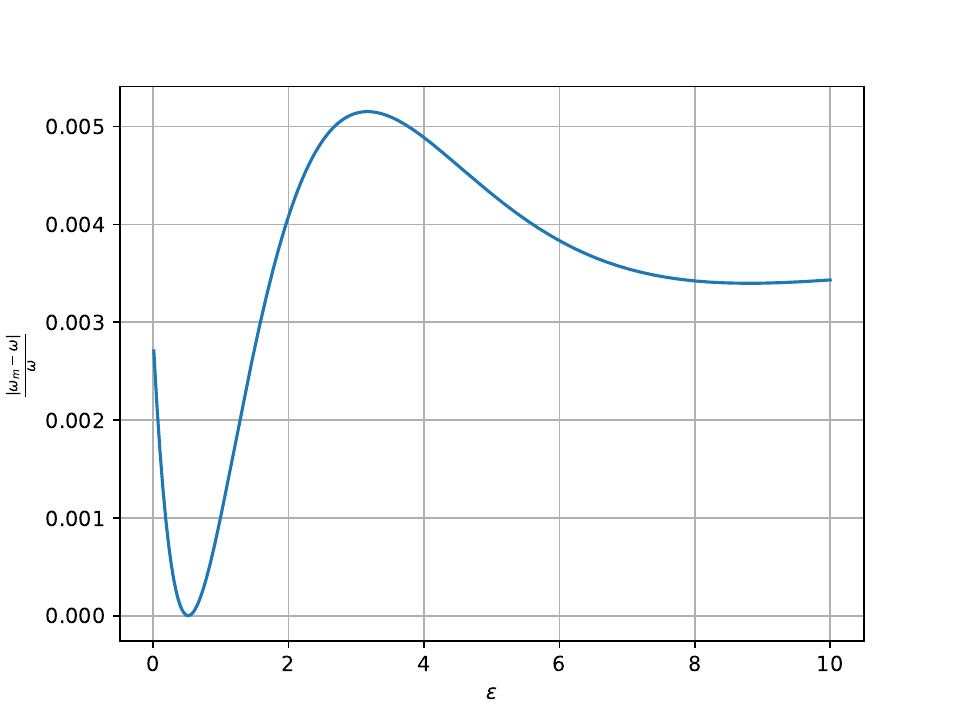}  
    }
    \caption{\textit{Left:} Updated comparison of the frequencies $\omega$, $\omega_m$, and $\omega_2$ as a function of the amplitude parameter $\varepsilon$. \textit{Right:} Relative error between the improved matched asymptotic solution and the numerically integrated solution which peaks at roughly 0.5\%.} \label{fig2}
\end{figure}

Consider a disk galaxy with mass density distribution given in Equation (\ref{eq:27}).
\begin{equation} \label{eq:27}
        \rho(r,z) = \rho_0 \exp{\left(-\frac{r}{R_g} - \frac{|z|}{z_g} \right)}
\end{equation}
In this scenario, $R_g$ is the scale radius of the disk and $z_g$ is the scale height of the disk as expected. By Equation (\ref{eq:6}), the natural frequency of simple harmonic motion near $z=0$ is $\omega_0^2 = 4\pi G \rho_0 \exp{\left(-\frac{r}{R_g}\right)}$. By the flatness approximation argued in Section \ref{sub:2.1}, we can solve for the effective vertical potential of our system using Equation (\ref{eq:28}).
\begin{equation} \label{eq:28}
    \frac{\partial^2\Phi}{\partial z^2} = \omega_0^2 e^{- \frac{|z|}{z_g}}
\end{equation}
This is a separable ODE with an analytic solution. The only physical solution that is twice-differentiable and symmetric about $z=0$ is given by Equation (\ref{eq:29}).
\begin{equation} \label{eq:29}
    \Phi_z = z_g^2 \omega_0^2 e^{-\frac{|z|}{z_g}} + z_g \omega_0^2 |z|
\end{equation}
This potential indeed becomes linear for $z \gg z_g$. The corresponding perturbation expansion for this potential computed using Equation (\ref{eq:16}) is given in Equation (\ref{eq:30}).
\begin{equation}\label{eq:30}
    \omega_2 = \omega_0 \left[1 - \frac{2}{3\pi}\varepsilon + \left(\frac{16}{27\pi^2} - \frac{1}{24}\right)\varepsilon^2 \right]
\end{equation}
For this potential, it is easy to compute that $g = z_g \omega_0^2$. Therefore, the large asymptotic frequency computed using Equation (\ref{eq:24}) is given by Equation (\ref{eq:31}).
\begin{equation} \label{eq:31}
    \omega_l = \omega_0 \sqrt{\frac{\pi^2}{8\varepsilon}}
\end{equation}
We now expand Equation (\ref{eq:26}) to second order in $\varepsilon$ and match the coefficients. We obtain the system of equations below (\ref{eq:32a}), (\ref{32b}), and (\ref{32c}).

\begin{equation}\label{eq:32a}
    1 = \frac{1}{\sqrt{A+B}}
\end{equation}
\begin{equation}\label{32b}
    -\frac{2}{3\pi} = \frac{BC - (8/\pi^2)}{2(A+B)^{3/2}}
\end{equation}
\begin{equation}\label{32c}
    \frac{16}{27\pi^2} - \frac{1}{24} = \frac{-2ABC^2 + B^2 C^2 - 6BC (8 / \pi^2) +3 \left(8/\pi^2\right)^2 }{8(A+B)^{5/2}}
\end{equation}
This system can be solved numerically to yield the values of the parameters $A \approx 0.2419$, $B \approx 0.7581$, and $C \approx 0.5093$. The matched asymptotic solution is then given by Equation (\ref{eq:33}).
\begin{equation}\label{eq:33}
    \omega_m(\varepsilon) \approx \omega_0 \left(0.2419 + 0.7581e^{-0.5093\varepsilon} + \frac{8z_g \omega_0^2 \varepsilon}{\pi^2 g}\right)^{-1/2}
\end{equation}
We can compare this to the true solution for $\omega_m$ using numerical integration. Recall that the frequency of the motion can be computed exactly using action-angle variables as was done in Section \ref{sub:2.3}. The relevant integral is given in Equation (\ref{eq:34}).
\begin{equation}\label{eq:34}
    \omega = 2\pi \left[\int_0^{z_0}\frac{4\sqrt{2}}{\sqrt{z_g^2 \omega_0^2 e^{-z_0/z_g} + z_g \omega_0^2 z_0 - z_g^2 \omega_0^2 e^{-|z|/z_g} - z_g \omega_0^2 |z|}} \, dz\right]^{-1}
\end{equation}
To make this numerically tenable, we work in length units of $z_g$ and time units of $\omega_0^{-1}$ such that $z_g = \omega_0 = 1$ and $\varepsilon = z_0$. We perform the numerical integration in Python using Simpson's rule with $10^5$ uniform steps. We also compute the relative error between $\omega$ and $\omega_m$, given by $\frac{|\omega_m - \omega|}{\omega} $, for comparison. The results of this numerical test are displayed in Figure \ref{fig1}. We observe that the two frequencies are almost equal over the range of $\varepsilon$. Unsurprisingly, the accuracy of the matched asymptotic model seems worse for $\varepsilon \sim 1$ where the solution is in its intermediate range. We have also plotted the frequency from Equation (\ref{eq:30}) to demonstrate the relatively short range over which the perturbative solution is viable and the large extent to which the matching improves the quality of the model. The primary source of error is likely truncation error due to only using three perturbation orders in the matching. This model performs relatively well, but we may attempt to improve it by experimenting with different forms of $f(\varepsilon)$ that satisfy the fundamental criteria we originally set. We  decrease the number of fitting parameters to reduce the possibility of overfitting to the first couple terms of the perturbation series. These seem to underestimate the true solution as shown in Figure \ref{fig1}. We consider removing the parameter $A$ in Equation (\ref{eq:26}) and performing the matching process again but only using the first and second order coefficients. We can manually add a constant term to ensure the solution goes to $\omega_0$ for $\varepsilon \rightarrow 0$. This results in the following system of equations (\ref{37}) and (\ref{38}).

\begin{equation} \label{37}
    -\frac{2}{3\pi} = \frac{\pi^2 AB - 8}{2\pi^2 A^{3/2}}
\end{equation}
\begin{equation}\label{38}
    \frac{16}{27\pi^2} - \frac{1}{24} = \frac{\pi^4A^2B^2 - 48\pi^2 AB + 192}{8\pi^4 A^{5/2}}
\end{equation}
Solving this yields $A \approx 0.9568$ and $B \approx 0.4320$ which means we add a small constant term of 0.0432 to constrain the limiting behavior. The resulting matched asymptotic solution is given in Equation (\ref{39}).
\begin{equation}\label{39}
    \omega_m \approx \left(0.0432 + 0.9568e^{-0.432\varepsilon} + \frac{8}{\pi^2}\varepsilon\right)^{-1/2}
\end{equation}
This introduces small errors in the Taylor expansion, but we demonstrate the improved results in Figure \ref{fig2}. The improvement is very apparent since the two curves are effectively on top of each other and the maximum error is reduced by roughly a factor of 4. Since the frequency is determined to high precision and our perturbative analysis dictates that the solution is sinusoidal, we can propose that the approximate solution for the vertical motion in an exponential disk is given by Equation (\ref{40}) where $\beta_0$ is some arbitrary phase shift determined by the initial conditions of the motion.
\begin{equation} \label{40}
    z(t) \simeq z_0 \cos{\left(\omega_m t + \beta_0\right)}
\end{equation}

\begin{figure}
    \centering
    \includegraphics[width=0.7\linewidth]{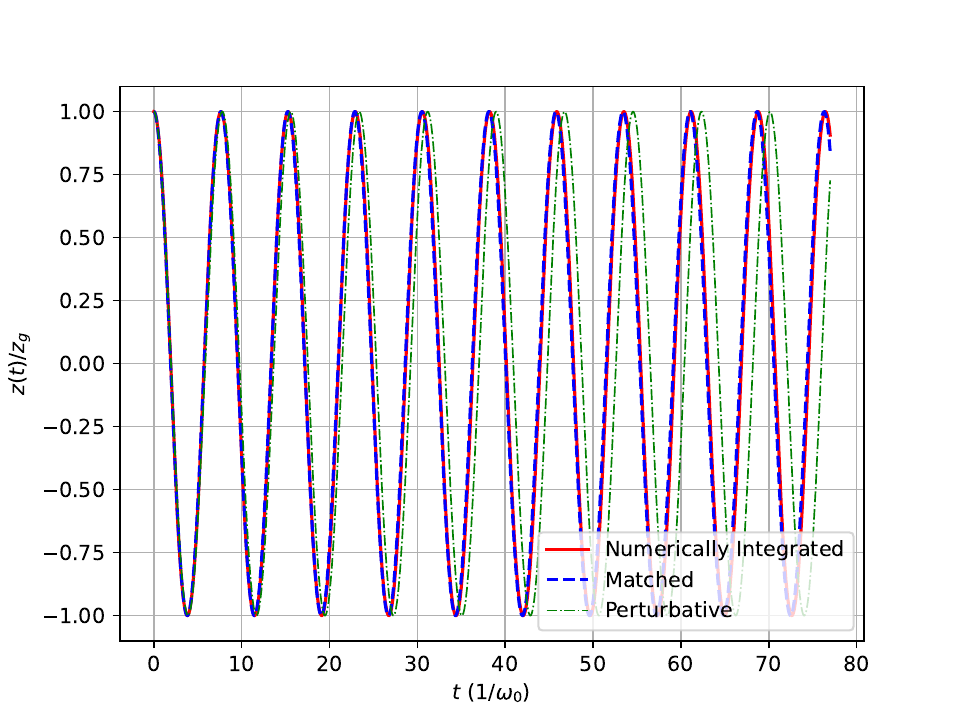}
    \caption{Comparison of $z(t)$ solved using numerical integration and the matched asymptotic model. The perturbative model of Equation (\ref{eq:30}) is also plotted for reference. The system is evolved over approximately 10 periods of the motion with the true and matched curves remaining effectively on top of each other.}
    \label{fig3}
\end{figure}

\noindent To test the accuracy of this approximate solution, we utilize a simple RK4 numerical integrator to solve the equations of motion directly from the potential in Equation (\ref{eq:29}). We set the energy of the system such that the amplitude of oscillations is $z_0 = z_g = 1$. The results of the numerical experiment are provided in Figure \ref{fig3}. The matched asymptotic model is very closely aligned with the true solution. Though the accuracy is generally determined by the value of $\varepsilon$, the two models drift out of phase after several hundred periods of the motion in general.

\subsection{The Isothermal Disk}
We apply the matched asymptotic method to one more example, the isothermal disk with vertical mass distribution given in Equation (\ref{41}) \citep{1942ApJ....95..329S}.

\begin{equation} \label{41}
    \rho(z) = \rho_0 \textrm{sech}^2{\left(\frac{z}{z_g} \right)}
\end{equation}
We follow the same procedure as before, noting that now $\omega_0^2 = 4\pi G \rho_0$. The resulting potential is given in Equation (\ref{42}) and the second order perturbation expansion is given in Equation (\ref{43}).

\begin{equation} \label{42}
    \Phi_z = \ln{\left[\cosh{\left(\frac{z}{z_g}\right)}\right]}
\end{equation}
\begin{equation} \label{43}
    \omega_2 = \omega_0\left(1 - \frac{\varepsilon^2}{8}\right)
\end{equation}
From this, we can repeat the process from Section \ref{sub:3.1} using the improved fitting method with only two parameters. The result of the matching is given in Equation (\ref{44}).

\begin{equation} \label{44}
    \omega_m \approx \left(-0.1154 + 1.1154e^{-0.7267\varepsilon} + \frac{8}{\pi^2}\varepsilon\right)^{-1/2}
\end{equation}

\begin{figure} 
    \centering
    
    \subfigure{
        \includegraphics[width=0.48\textwidth]{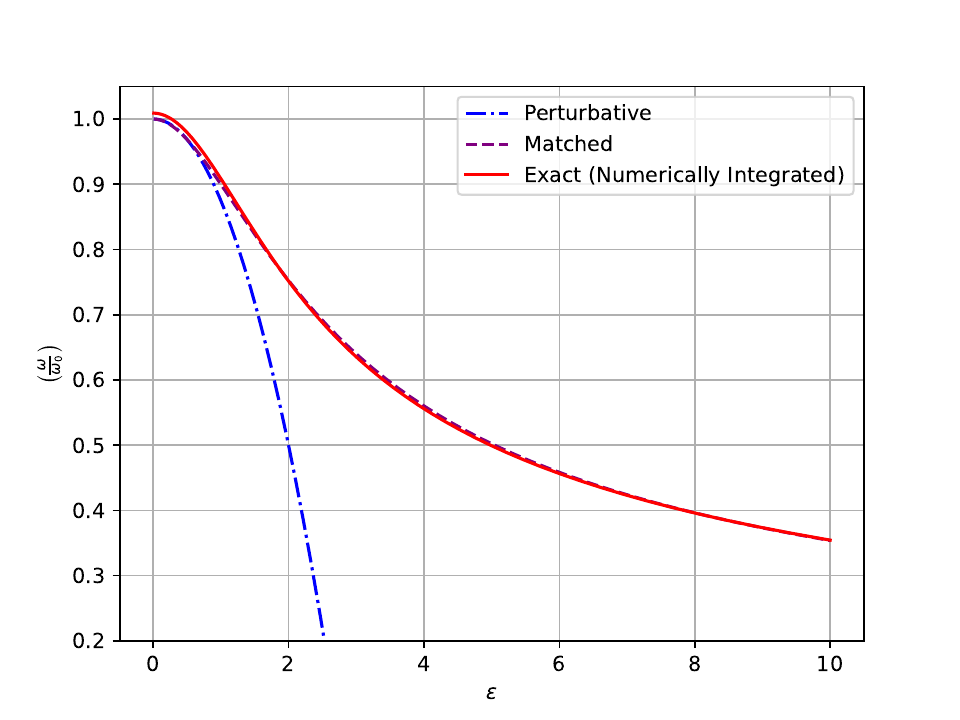}  
    }
    \subfigure{
        \includegraphics[width=0.48\textwidth]{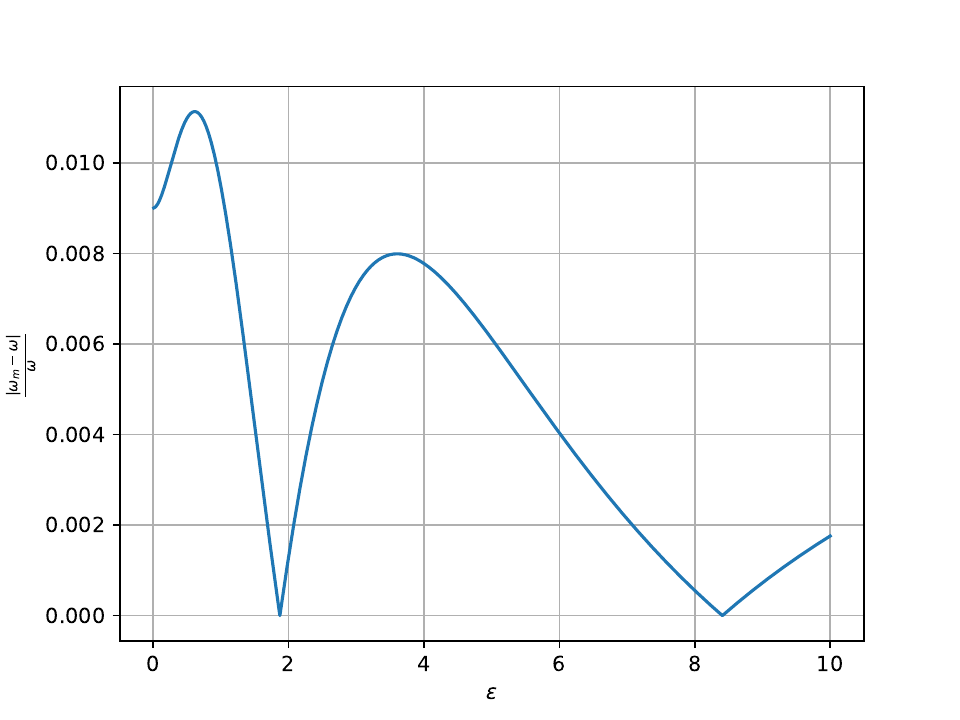}  
    }
    \caption{\textit{Left:} Updated comparison of the frequencies $\omega$, $\omega_m$, and $\omega_2$ as a function of the amplitude parameter $\varepsilon$ for the isothermal disk. \textit{Right:} Relative error between the improved matched asymptotic solution and the numerically integrated solution which peaks at roughly 1.0\%.} \label{fig4}
\end{figure}

\begin{figure}
    \centering
    \includegraphics[width=0.7\linewidth]{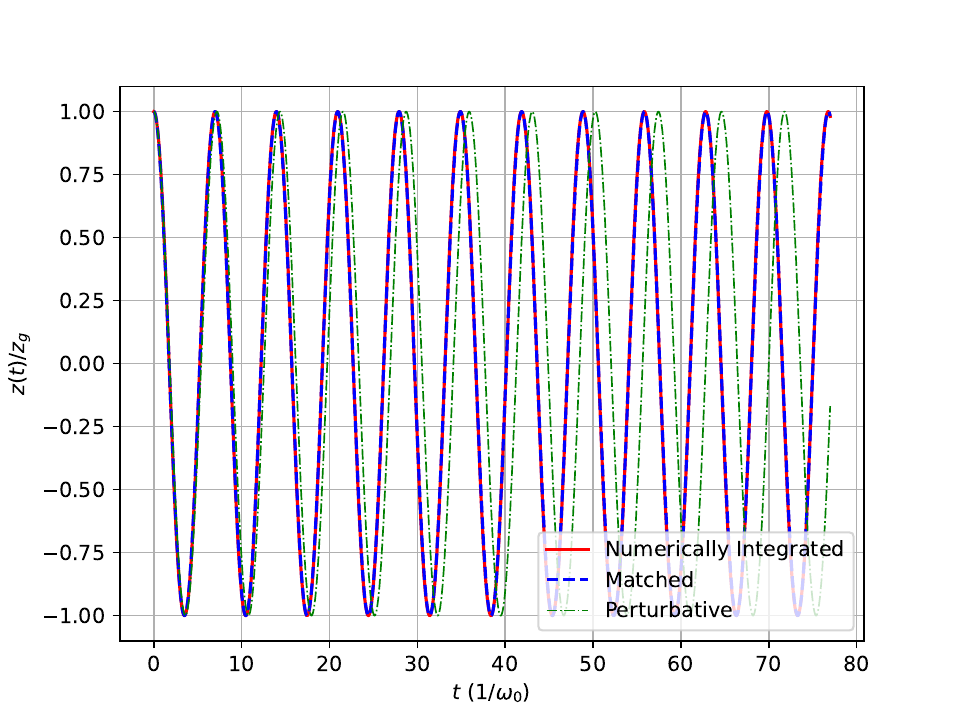}
    \caption{Comparison of $z(t)$ solved using numerical integration and the matched asymptotic model for the isothermal disk. The perturbative model of Equation (\ref{43}) is also plotted for reference. The system is evolved over approximately 10 periods of the motion with the true and matched curves again aligning very closely.}
    \label{fig5}
\end{figure}

\noindent We can evaluate the accuracy of this model using the same numerical integration procedure as before but with the potential from Equation (\ref{42}). The results are presented in Figure \ref{fig4}. We achieve accuracy to within a few percent as we had in the previous example. The general matched asymptotic solution is again given by a sinusoid of the form in Equation (\ref{40}). We numerically test the new potential using the same RK4 setup and with the same initial conditions. The results of the numerical comparison are given in Figure \ref{fig5}.


\section{Discussion}\label{sec:discussion}
Based on our numerical results, the matched asymptotic method we derived seems applicable to general thin disk structures. The two models we elected to apply our method to are also two of the more ubiquitous models for galactic disk structure, so equations (\ref{39}) and (\ref{44}) are potentially the most useful results of this method. In applying this method, we note that the results depend on the choice of $f(\varepsilon)$ as evidenced by our discussion in Section \ref{sub:2.4}. The most reasonable choice of basis functions for this type of problem seems to be exponential functions which reflect the rapidly decaying vertical density of most galactic disks.

The model we formulated cannot accurately describe the vertical motion when the point mass undergoes significant radial shifts in its orbit. This is because the oscillation frequency $\omega_0$ is a function of the local density and hence the radius as well. This means the parameter $\omega_0$ in our governing Hamiltonian is time-dependent which implies that energy is not conserved for $\mathcal{H}_z$. However, if $\omega_0$ varies slowly in comparison to the oscillation time scale, then the action of the system $J$, i.e. the volume enclosed by the orbit in phase space, will be invariant during this change \citep{goldstein1980classical}. A more mathematical description of this condition is given in Equation (\ref{alpha}).

\begin{equation} \label{alpha}
    |\alpha| \ll \omega, \quad \alpha \equiv \frac{\dot{\omega}(t)}{\omega(t)}
\end{equation}
We may then attempt to model how the oscillation amplitude varies with this slow change in the system's energy. Combined with our frequency model from the previous sections, this would give us a more complete description of the dynamics of the system. In the regime of $\varepsilon \ll 1$, the system behaves as a simple harmonic oscillator which has action $J_s = \pi z_s(t)^2 \omega_0(t)$ where $z_s(t)$ is the time varying amplitude for the simple harmonic oscillator and $\omega_0(t)$ is the time varying natural frequency of vertical oscillations. By adiabatic invariance, we have the relation in Equation (\ref{46}) where $z_0$ is the amplitude at the initial time.
\begin{equation} \label{46}
    z_s(t) = z_0 \sqrt{\frac{\omega_0(0)}{\omega_0(t)}}
\end{equation}
The amplitude evidently scales as $\omega_0(t)^{-1/2}$. Therefore, since $\omega_0$ should only depend on the local density, the scaling of the vertical oscillation amplitude is completely determined by the position of the object in its orbit. For the case of large $\varepsilon \gg 1$ where the potential becomes linear, the same calculation can be performed to yield the relation in Equation (\ref{47}) where $z_l(t)$ is the time varying amplitude for the linear oscillator. 
\begin{equation} \label{47}
    z_l(t) = z_0\left(\frac{\omega_0(0)}{\omega_0(t)}\right)^{2/3}
\end{equation}
Based on these results, the matched frequency $\omega_m$ can be computed based on the position and initial amplitude alone, assuming the condition in Equation (\ref{alpha}) holds. The choice of scaling relation naturally depends on the initial amplitude $z_0$ with equations (\ref{46}) and (\ref{47}) being suitable for $\varepsilon \lesssim 1$ and $\varepsilon \gtrsim 1$ respectively. The adiabatic invariance also implies that the model may be generalized to some structures that are not axisymmetric if the density in the plane of the disk varies slowly. However, for structures like spiral arms, it is likely that the adiabatic condition is violated due to resonant scattering \citep{2008ApJ...684L..79R, 2020ApJ...904..137T}. Under the reasonable assumption that the local density of the galactic disk increases for decreasing radius $r$, these results also suggest that the vertical oscillation amplitude decreases as the object is pulled inward.
\section{Conclusions}\label{conc}
We present an analytical model for the non-planar dynamics of a point mass perpendicular to a thin, axisymmetric galactic disk. We describe a general method by which the vertical oscillation frequency of a point mass may be calculated using perturbation theory and asymptotic analysis. Notably, we compute the relationship between the vertical displacement of the point mass and its characteristic oscillation frequency. We apply the model to two galactic mass distributions and maintain reasonable accuracy, which is evidenced by numerical simulations. The model also describes the relationship between the radial and vertical motion and offers a more complete description of non-planar orbits applicable to a wider range of scenarios.

This model presents the greatest utility for numerical galactic dynamics where the dynamics of potentially millions of astronomical objects \citep{2015MNRAS.450.4070W} need to be computed accurately and inexpensively. An analytical model for the vertical motion that completely bypasses numerical integration of the equations of motion in $z$ presents an opportunity to design more computationally efficient $N$-body simulations. Since we have shown that the $z$-dynamics can be computed directly from the radial dynamics under our assumptions, only the equations of motion of the remaining spatial dimensions would need to be numerically integrated. 

The model may also have applications in the study of compact objects, particularly in the study of the decay of massive black hole (MBH) binaries. In the current framework of MBH binary decay outlined by \cite{1980Natur.287..307B}, the primary decay mechanism at ranges of $\sim$1 kpc is dynamical friction. In general, the strength of dynamical friction depends on the local density of gas and stars \citep{1943ApJ....97..255C, 1999ApJ...513..252O, 2012ApJ...745...83A}. In most models for the vertical structure of a galaxy, the density decreases exponentially with height which presents a challenge to simulations which focus exclusively on dynamics in the plane of the disk \citep{2020ApJ...896..113L, Qian_2024}. Even at small vertical displacements, the strength of dynamical friction can oscillate significantly due to vertical oscillations which planar simulations cannot account for. With our model, the characteristic frequency of these oscillations can be accurately calculated without requiring the implementation of another spatial dimension to these types of simulations. 

\begin{acknowledgements}
\noindent S.G. acknowledges the support of David Ballantyne, Tamara Bogdanovic, Roman Grigoriev, Gongjie Li, and Steinn Sigurdsson in providing valuable feedback and the necessary resources to conduct this research.
\end{acknowledgements}

\bibliography{Asymptotic_Matching_Preprint_122124}{}
\bibliographystyle{aasjournal}



\end{document}